\pdfoutput=1

\documentclass[10pt,letterpaper]{article}

\usepackage{ccn}
\usepackage{pslatex}
\usepackage{apacite}

\usepackage{url}
\usepackage{algorithm}
\usepackage{algorithmic}
\usepackage{graphicx}
\usepackage{amsthm}
\usepackage{amssymb}
\usepackage{multicol}
\usepackage{amsmath}
\usepackage{caption}

\title{Visualizing Representational Dynamics with Multidimensional Scaling Alignment}
 
\author{{\large \bf Baihan Lin and Nikolaus Kriegeskorte$^*$} \\
  \{Baihan.Lin, N.Kriegeskorte\}@columbia.edu \\
Zuckerman Mind Brain Behavior Institute, Columbia University, New York, NY, USA 10027 \\
$^*$corresponding author
  \AND {\large \bf Marieke Mur} \\
  MMur@uwo.ca\\
The Brain and Mind Institute, University of Western Ontario, London, ON, Canada N6A 5B7 \\
  \AND {\large \bf Tim Kietzmann} \\
  Tim.Kietzmann@mrc-cbu.cam.ac.uk\\
MRC Cognition and Brain Sciences Unit, Cambridge University, Cambridge, United Kingdom CB2 7EF
}

\begin{document}

\maketitle

\section{Abstract}
{\bf

Representational similarity analysis (RSA) has been shown to be an effective framework to characterize brain-activity profiles and deep neural network activations as representational geometry by computing the pairwise distances of the response patterns as a representational dissimilarity matrix (RDM). However, how to properly analyze and visualize the representational geometry as dynamics over the time course from stimulus onset to offset is not well understood. In this work, we formulated the pipeline to understand representational dynamics with RDM movies and Procrustes-aligned Multidimensional Scaling (pMDS), and applied it to neural recording of monkey IT cortex. Our results suggest that the the multidimensional scaling alignment can genuinely capture the dynamics of the category-specific representation spaces with multiple visualization possibilities, and that object categorization may be hierarchical, multi-staged, and oscillatory (or recurrent).

}

\begin{quote}
\small
\textbf{Keywords: } MDS, RSA, GPA, Neuroimaging
\end{quote}

\section{Introduction}

% \begin{figure}[tb]
% \centering
%     \includegraphics[width=0.8\linewidth]{./RepDyn.jpg}\par\caption{What, where and when \cite{mur2014s}.}\label{fig:repDyn}
% \end{figure}

In recent years, technological innovations in computer vision have produced biologically-plausible models for human visual information processing. Among these models are goal-driven deep feedforward hierarchical neural networks, which have been proposed to model the ventral stream of visual cortex, the ``what pathway'' in the brain thought to underlie object recognition \cite{yamins2016using,khaligh2014deep,kriegeskorte2015deep}. However, there is a discrepancy in the hierarchical depth between the primate ventral visual stream ($<$ 10 stages of representation) and state-of-the-art computer-vision models ($>$ 100 layers). The primate visual system might make up for its limited hierarchical depth by recycling its resources through time, via recurrent connections and attention mechanisms, all of which require the analysis of the entire time-series of the dynamics in the brain. Few studies have investigated the dynamics of visual perception (for instance, object identity and categorization) and their representational changes \cite{hung2005fast,freiwald2010functional}. 

The scarcity of methods to characterize the representational dynamics creates a major barrier to answer interesting questions such as: how are objects represented in the brain over the time course from early perception to categorical decision making, does the object identification or visual categorization follows a hierarchical classification paradigm; do different classes of objects merge and branch at different time points based on different tasks or recurrence paradigm; are these representational dynamics oscillatory or recurrent? 
% (Figure \ref{fig:repDyn}) 

A central challenge is exactly to test computational theories implemented in deep neural network models with exactly this type of time-stamped brain-activity data. Analyses of representational geometry can help us to compare representations between biological brains and computational models, and to understand brain computation as the transformation of representational similarity structure across stages of processing \cite{kriegeskorte2013representational}. Representational Similarity Analysis (RSA) uses a region's representational dissimilarity matrix (RDM) as a multivariate summary statistic that characterizes the representational geometry using metric distances in representational space. This is useful to obviate the need for a detailed point-to-point mapping between neurons in the brain and units of a model. 

Traditional RSA usually considers the entire time series of the neural measurement as the response pattern. In that sense, the dynamics is collapsed into one data point to characterize the brain-activity corresponding to a specific stimulus. However, it is unclear how to properly capture the representational dynamics, i.e., how the representations evolve over time. In this study, we propose a framework to first extract the snapshots of representational spaces with sliding-window RSA, and then align each frame of this RDM ``movie'' (as snapshots of the representational space) with Generalized Procrustes Analysis (GPA). We presented the visualizations on the data of monkey's stimulus-driven single-electrode recording, and demonstrated several neuroscience insights on visual object categorization revealed from the proposed method.

\section{Method}

\subsection{Representational Similarity Analysis (RSA)}

Representational Similarity Analysis (RSA) characterizes internal representations of a brain network by estimating all pairwise distances (as the representational geometry) across a large set of input conditions. These representational geometries are invariant to rotations of the underlying high-dimensional activation space. RSA involves two steps: (1) compute the dissimilarity for each pair of stimuli; (2) correlate RDMs to assess to what extent the brain representation reflects stimulus properties, can be accounted for by different computational models, and is reflected in behaviours. 

\subsection{Multidimensional Scaling (MDS)}

As a popular non-linear dimensionality reduction method, Multidimensional Scaling (MDS) rearrange the location of a set of data points from a set, given a distance matrix characterizing the distances between each pair of objects in a set (for instance, the RDM computed from the response patterns of different stimuli), into an N-dimensional space such that the between-object distances are preserved \cite{buja2008data}. 

\subsection{Generalized Procrustes Analysis (GPA)}

In computer vision and signal processing, Procrustes analysis is usually used to analyse the statistical distribution of a set of shapes. The Generalized Procrustes analysis (GPA) compares three or more shapes to an optimally determined "mean shape" and can align all the shapes according to this mean shape as a reference frame \cite{gower1975generalized}. GPA solves the mean shape iteratively by optimizing against the Procrustes distance, a metric to minimize in order to superimpose any pair of shape or time frame instances annotated by landmark points. The analysis starts from choosing an arbitrary reference frame, and then superimpose all instances to current reference shape. If Procrustes distance between the reference shape and the computed mean shape of the current set of superimposed shapes is above a certain threshold, the reference shape is set to the mean shape to continue above steps iteratively, until the Procrustes distance between the two is small enough within the trivial threshold.

\subsection{Procrustes-aligned MDS (pMDS)}

In our specific problem, the RDM movie consists of the representational shapes of each time point without any intertemporal information. We apply GPA to the MDS embeddings computed from RDMs at each time point, such that each frame are optimally aligned to all other time frames. The Procrustes analysis has the option to constrain rotation, scaling and reflection, but in our case, we only allow the rotation and reflection, because the scaling contains information about the how the representations diverge and converge over time. Because Procrustes alignment doesn't distort the geometrical information between each stimuli (constrained by the individual RDM at each time point), the Procrustes-aligned MDS (pMDS) can offer us a genuine and illustrative visualization of the representational dynamics over time.

\section{Results}
\label{sec:results}

\subsection{Data of neural population code}

The data used to demonstrate the proposed method are the monkey single-electrode recordings from the inferior bank of the ST segment \cite{bell2011relationship}. Two adult male rhesus monkeys were shown 100 grayscale object images from five different categories each with 20 instances (faces, fruits, places, body parts and objects) in a serial visual presentation. RSA was further applied to select visually-responsive neurons and extract single-trial response patterns from spike-density function. The recordings were truncated into sections of 821 ms (starting from 100ms before stimulus onset). RDM movies were generated using a sliding window of 21 ms with cross-validated spike rate distance (SRD) as the reponse-pattern dissimilarity measure. 

\subsection{MDS alignment reveals smooth transition over time}

\begin{figure}[tb]
\centering
    \includegraphics[width=1\linewidth]{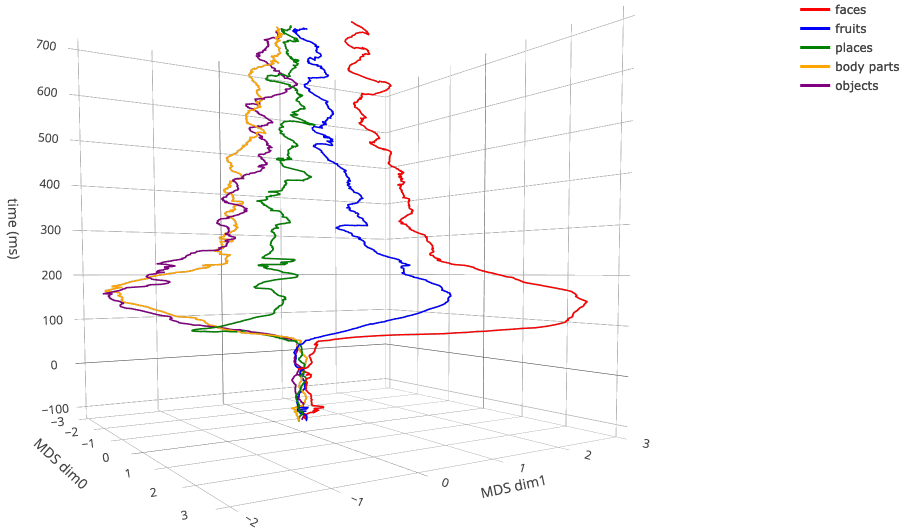}\par\caption{3D plot of Procrustes-aligned MDS over time.}\label{fig:3d}
\end{figure}

We performed GPA on the MDS embeddings computed from each time frame of RDM movies MDS based on the stimulus label (not the category label). As shown in Figure \ref{fig:3d}, the average trajectory of all the data points of the same category are plotted in the MDS space over time. The dynamics over time can be reasonably visualized while the separation of each categories (as between-category distances) is well preserved. From the 3D plot of the representational space, the separation of each categories happens around 80ms and reached a maximum distinction at around 150ms after stimulus onset, then the trajectories gradually converges over time in an oscillatory fashion after the stimulus offset (at 300ms).  A movie of Procrustes-aligned MDS plots is also generated and can be accessed at \url{https://youtu.be/WQbgDCq7Dhg}, where each data point (a stimulus instance) can be distinctively tracked as moving seamlessly across each time frame. 

\subsection{Hierarchical visual categorization with major stages}

\begin{figure}[tb]
\centering
    \includegraphics[width=1\linewidth]{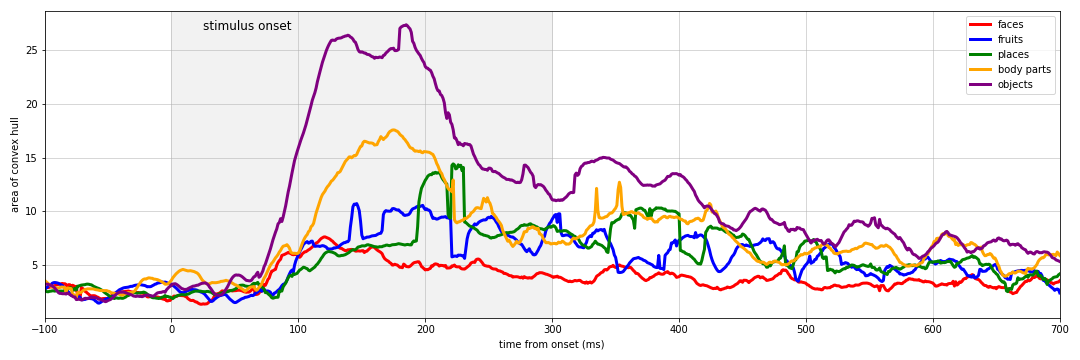}\par\caption{Changing areas of convex hulls over time.}\label{fig:2d_convex_hulls}
\end{figure}

\begin{figure*}[tb]
\centering
    \begin{minipage}{.3\textwidth}\centering
    \includegraphics[width=0.83\linewidth]{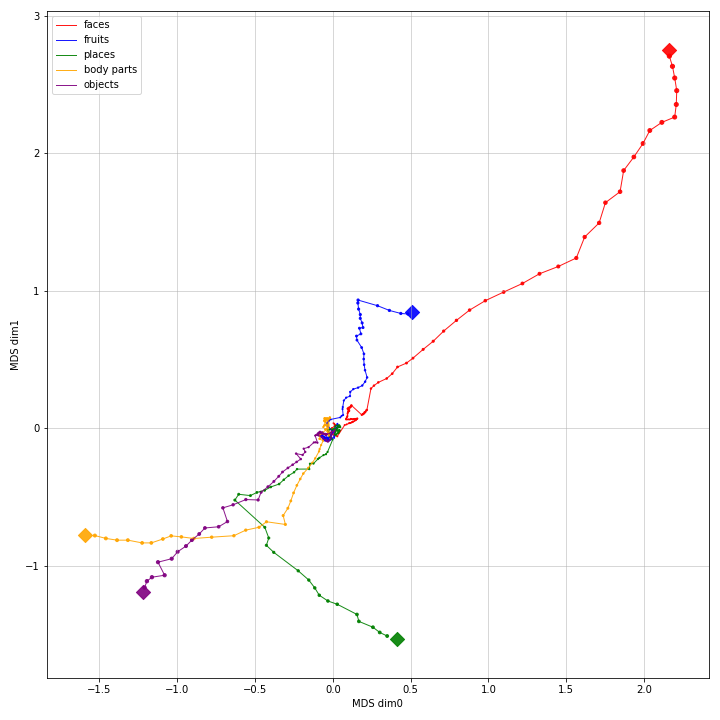}\par\caption{During onset (0-100ms).}\label{fig:2d_0_100}
    \end{minipage}
    \begin{minipage}{.3\textwidth}\centering
    \includegraphics[width=0.83\linewidth]{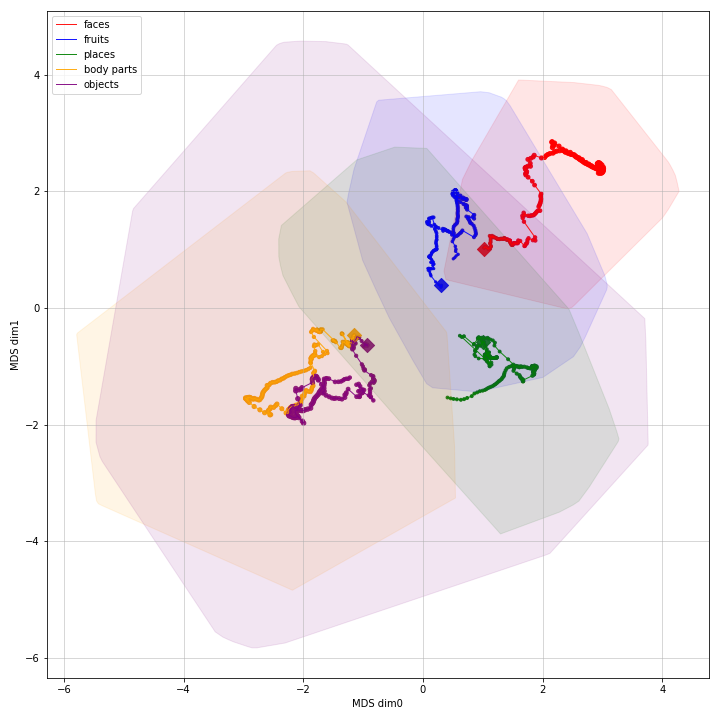}\par\caption{During onset (100-300ms).}\label{fig:2d_100_300}
    \end{minipage}
    \begin{minipage}{.3\textwidth}\centering
    \includegraphics[width=0.83\linewidth]{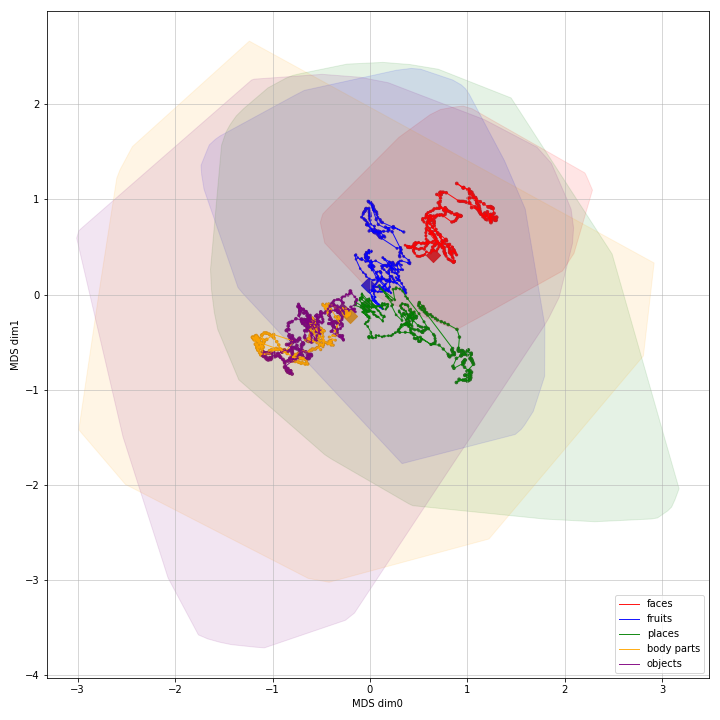}\par\caption{After offset (300-800ms).}\label{fig:2d_300_800}
    \end{minipage}
% \caption*{The trajectories of the average representation of each category over time (dot size as the standard deviations across all stimuli in the category, and square as the end of each trajectory, convex hulls created by all stimuli within the time range and category).}
\end{figure*}

Given the intuition from the 3D visualization, we further explored segments of the representational dynamics. Figure \ref{fig:2d_0_100} demonstrated the average representational trajectories of each categories in the first 100ms after stimlus onset (with the end marked as square and dot size indicating the standard deviation across different stimulus instances). We see that the categories faces and fruits diverges become discriminable rapidly in the IT population code due to their distinct visual dissimilarity, while places, objects and body parts diverge much later in time. Among the three late classes, the separation of the objects and body parts happens even later in time, suggesting a hierarchical process of categorization. Later during the stimulus is on, the representations of each category seems to be dwelling around their own cluster in a slowly drifting fashion, as shown in Figure \ref{fig:2d_300_800} (the convex hulls are plotted for all stimuli within the time range in the selected category). After the stimulus offset, the average trajectory of each category gradually converge into proximity, as shown in Figure \ref{fig:2d_300_800}, where the convex hulls of each categories gradually merge into one. The representational space for each category (as indicated by the areas of the convex hull covering all stimulus instances within the category) also follows several major segments (Figure \ref{fig:2d_convex_hulls}): a peak around 100 to 200 ms after onset and another bump after the stimlus offset around 300 to 400 ms. Further investigations can potentially illuminate the role of working memory and other factors that might contribute to the second rise of the representational areas.

\subsection{Temporal analysis of aligned representations}

\begin{figure}[tb]
\centering
    \includegraphics[width=0.9\linewidth]{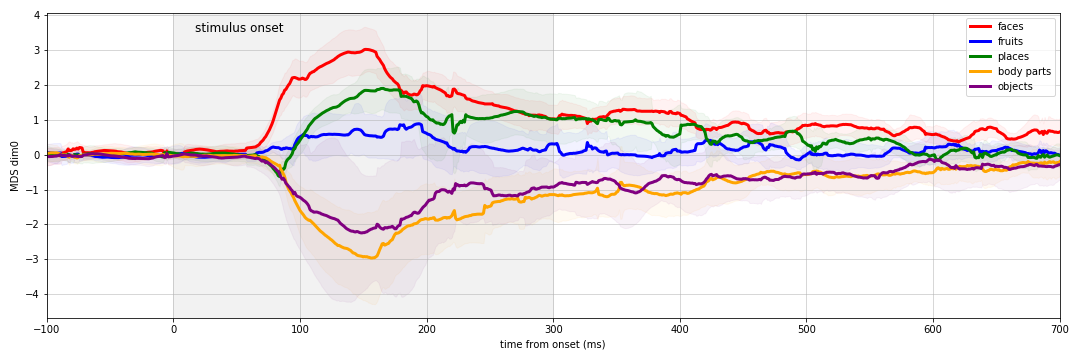}
    \includegraphics[width=0.9\linewidth]{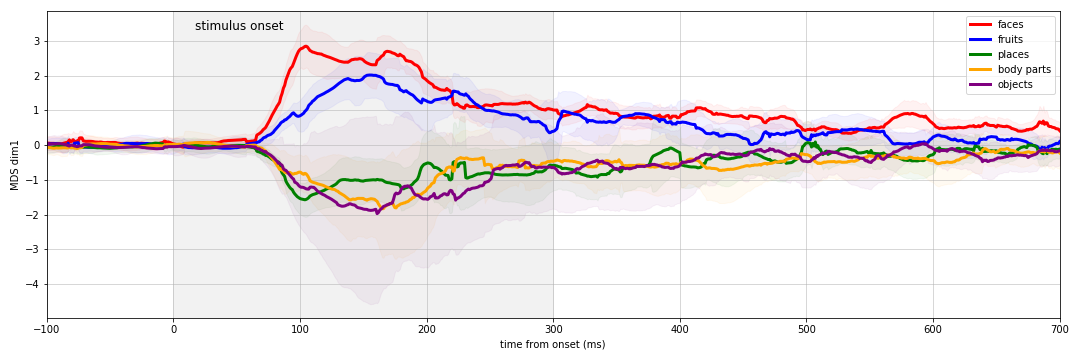}
    \includegraphics[width=0.9\linewidth]{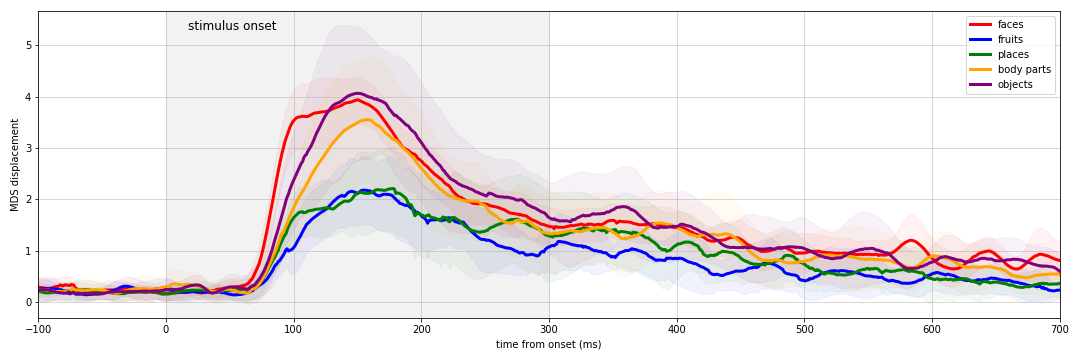}
    \includegraphics[width=0.9\linewidth]{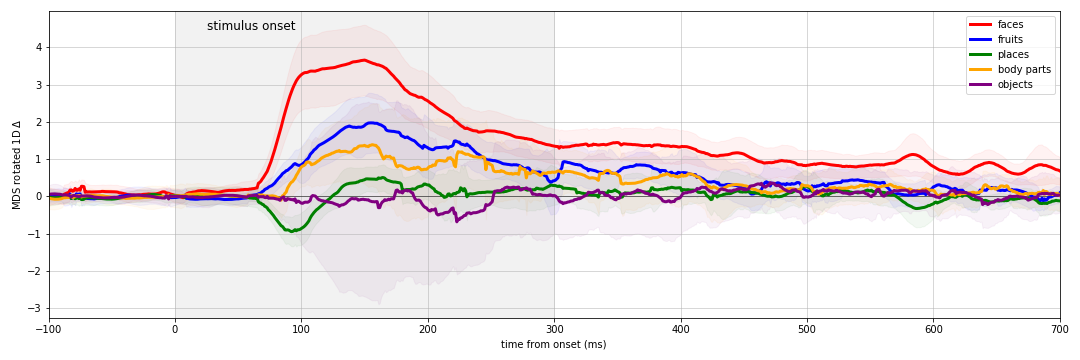}
    \par\caption{MDS embeddings over time (dim0, dim1, the displacement from origin, maximum incremental changes).}\label{fig:mds_2d}
\end{figure}

With the aligned representations, other temporal analyses can be applied to compare between time points. Figure \ref{fig:mds_2d} offers a subset of such inquires that can potentially offer neuroscience insights. For instance, the MDS displacement away from the origin (the third plot) indicates a similar dual bump feature during stimulus onset and offset, suggesting the drifting of the centroids of each categorical representation follows a unique dynamics that was not previously widely understood. In the fourth plot, we aligned the signal for each category onto the category-specific 1D plane where the signal amplitudes are maximized. Initial analysis reveals that there seems to be interesting oscillations worth further investigation. Due to GPA's preservation of geometrical information of the pairwise stimulus-specific response pattern, these temporal analyses offer none artificial insights on the variance or noise (due to the scaling restriction) and directionality (due to the minimized Procrustes distance).

\section{Conclusion and Future Work}

% \clearpage

We here proposed a representation alignment method to extend the RSA framework to analyze time-stamped brain-activity profiles as representational dynamics. From the neural data, RDM movies are computed with as sliding-window snapshots of representational geometry, and then aligned across all time points with generalized Procrustes analysis. We applied the proposed method to the single-electrode recording of monkey's IT cortex viewing 100 images of 5 categories. The results demonstrated that the alignment can reasonably capture the temporal dynamics of the representation space for each category, and reveal insights on the hierarchical separations of classes and possibly connection with other mechanisms such as working memory and oscillatory behaviors. 

\begin{figure}[tb]
\centering
    \begin{minipage}{.45\linewidth}
    \includegraphics[width=0.95\linewidth]{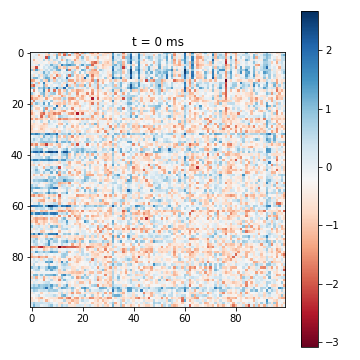}\end{minipage}
    \begin{minipage}{.45\linewidth}
    \includegraphics[width=0.95\linewidth]{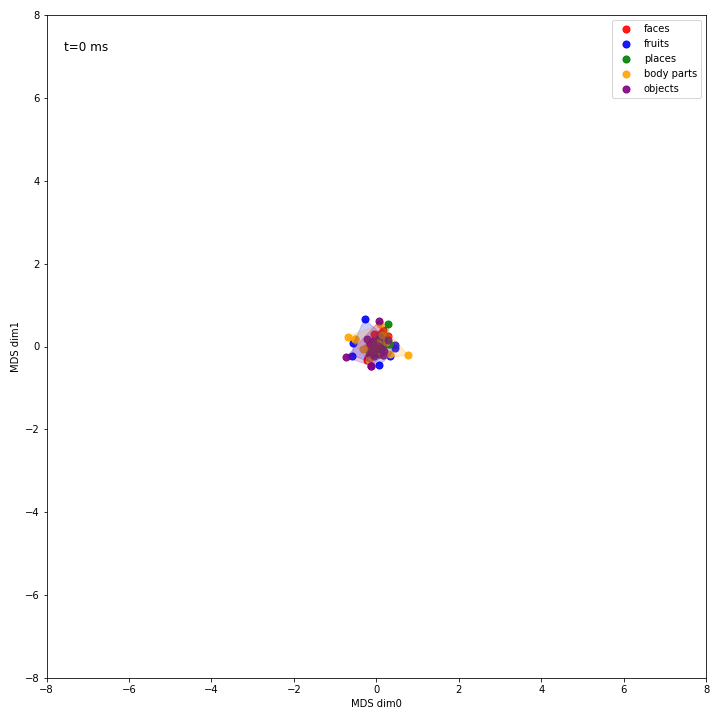}\end{minipage}
    \begin{minipage}{.45\linewidth}
    \includegraphics[width=0.95\linewidth]{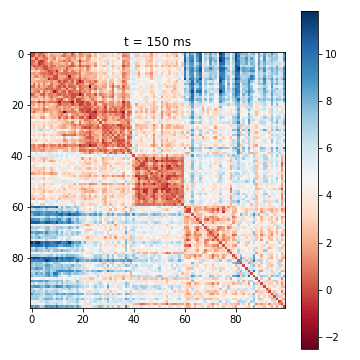}\end{minipage}
    \begin{minipage}{.45\linewidth}
    \includegraphics[width=0.95\linewidth]{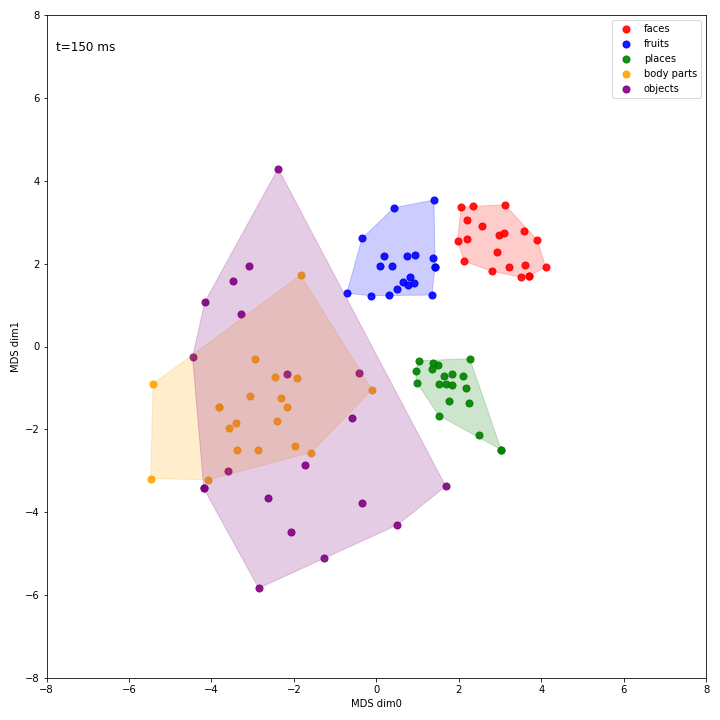}\end{minipage}
    \begin{minipage}{.45\linewidth}
    \includegraphics[width=0.95\linewidth]{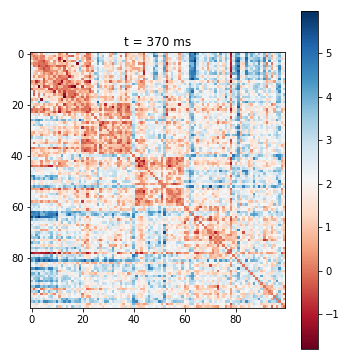}\end{minipage}
    \begin{minipage}{.45\linewidth}
    \includegraphics[width=0.95\linewidth]{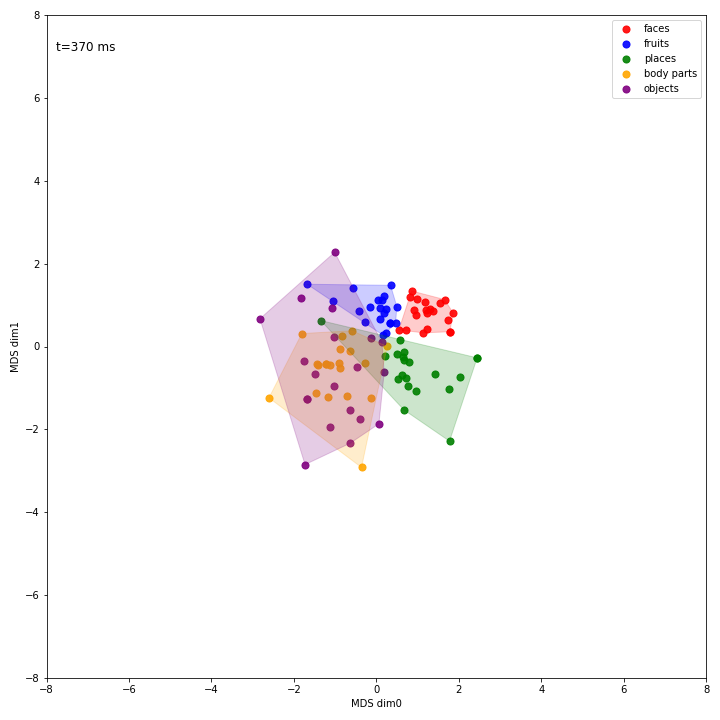}\end{minipage}
    % \begin{minipage}{.4\linewidth}
    % \includegraphics[width=0.95\linewidth]{./RDM_s1_t_500.png}\end{minipage}
    % \begin{minipage}{.4\linewidth}
    % \includegraphics[width=0.8\linewidth]{./stm100_s1_cMDS_t_600.png}\end{minipage}
    \caption{RDM and MDS at example time points (onset at 0ms, during onset at 150ms, and after offset at 370ms).}\label{fig:2d_rdm_mds}
\end{figure}

Other than working with RDM movies, there are several alternative methods to study representational dynamics. One such approach under ongoing investigation is to work with raw data of the time-series measurements, by directly extracting the pairwise intertemporal distances into a full RDM of dimension $(N\times T)\times (N\times T)$ with N as number of stimuli and T as number of time points. However, there are clear advantages of our currently proposed multidimensional scaling alignment of RDM movie over this alternative approach of RSA of full RDM: (1) the algorithmic complexity of working with the full RDM is so expensive that it's computationally prohibitory, while Procrustes-aligned MDS is very scalable and light-weighted; (2) the pairwise distances of time series segment is an ongoing challenge and topic of interests in text mining and bioinformatics, that requires deeper understanding to apply in a logical way. Another alternative is to simply use the snapshots RDMs themselves. Figure \ref{fig:2d_rdm_mds} compares these two methods, where RDM's grid-like and MDS's patch-like visualizations each offer a unique insight of the pattern. Future and ongoing work includes the application of this method to the neuroimaging data of different brain regions and time scales to explore whether the representations are also recurrent as the neural recordings \cite{kietzmann2019recurrence}, as well as visualizing the representational dynamics of deep neural networks to understand their behaviors. 

% \begin{figure*}[tb]
% \centering
%     \begin{minipage}{.24\textwidth}
%     \includegraphics[width=0.95\linewidth]{./RDM_s1_t_0.png}\end{minipage}
%     \begin{minipage}{.24\textwidth}
%     \includegraphics[width=0.95\linewidth]{./RDM_s1_t_150.png}\end{minipage}
%     \begin{minipage}{.24\textwidth}
%     \includegraphics[width=0.95\linewidth]{./RDM_s1_t_370.png}\end{minipage}
%     \begin{minipage}{.24\textwidth}
%     \includegraphics[width=0.95\linewidth]{./RDM_s1_t_500.png}\end{minipage}
%     \begin{minipage}{.24\textwidth}\lefting
%     \includegraphics[width=0.8\linewidth]{./stm100_s1_cMDS_t_100.png}\end{minipage}
%     \begin{minipage}{.24\textwidth}\lefting
%     \includegraphics[width=0.8\linewidth]{./stm100_s1_cMDS_t_250.png}\end{minipage}
%     \begin{minipage}{.24\textwidth}\lefting
%     \includegraphics[width=0.8\linewidth]{./stm100_s1_cMDS_t_470.png}\end{minipage}
%     \begin{minipage}{.24\textwidth}\lefting
%     \includegraphics[width=0.8\linewidth]{./stm100_s1_cMDS_t_600.png}\end{minipage}
%     \caption{RDM and MDS at example time points (during onset at 0 and 150ms and after offset at 370 and 500ms).}\label{fig:2d_rdm_mds}
% \end{figure*}

% \clearpage
\bibliographystyle{apacite}
\bibliography{main}

\setlength{\bibleftmargin}{.125in}
\setlength{\bibindent}{-\bibleftmargin}

\end{document}